\documentclass[preprint,12pt]{elsarticle}

\usepackage{graphicx}  
\usepackage{graphics}
\usepackage{dcolumn}   
\usepackage{bm}        
\usepackage{amssymb}   
\usepackage{color}
\usepackage{amsmath}

\usepackage[font=scriptsize]{caption}

\journal{Physica A}

\begin{document}

\begin{frontmatter}



\title{The evolution of network controllability in growing networks}

\author[label1]{Rui Zhang}
\author[label1]{Xiaomeng Wang}
\author[label2]{Ming Cheng}
\author[label1]{Tao Jia}
\ead{tjia@swu.edu.cn}
\address[label1]{College of Computer and Information Science, Southwest University, Chongqing, 400715, P. R. China}
\address[label2]{School of Rail Transportation, Soochow University, Suzhou, Jiangsu, 215131, P. R. China}

\begin{abstract}
The study of network structural controllability focuses on the minimum number of driver nodes needed to control a whole network. Despite intensive studies on this topic, most of them consider static networks only. It is well-known, however, that real networks are growing, with new nodes and links added to the system. Here, we analyze controllability of evolving networks and propose a general rule for the change of driver nodes. We further apply the rule to solve the problem of network augmentation subject to the controllability constraint. The findings fill a gap in our understanding of network controllability and shed light on controllability of real systems.
\end{abstract}


\end{frontmatter}

\section{Introduction}
How to control a complex system is one of the most challenging problems in science and engineering with a long history. During recent years, there were a significant amount of works addressing the controllability of complex networks, our ability to drive a network from any initial state to any desired final state within finite time\cite{kalman1963mathematical, luenberger1979introduction, chui2012linear}. A general framework based on structural controllability of linear systems was first proposed to identify the minimum set of driver nodes (MDS) \cite{liu2011controllability}, whose control leads to the control of the whole network. Following it, related problems under this framework were also investigated, ranging from the cost of control \cite{yan2012controlling, Yan2015Spectrum, sun2017closed} to the robustness and optimal of controllability \cite{chen2017robustness, ding2016optimizing, wang2017physical, nie2014robustness}, from the multiplicity feature in control \cite{jia2013emergence, jia2013control, jia2014connecting, zhang2017efficient} to the controllability in multi-layer or temporal networks \cite{posfai2014structural, posfai2016controllability, pan2014towards, menichetti2016control}, and more \cite{gao2014target, tang2014synchronization, zhao2015intrinsic, wang2017spectral}. This framework was also applied to different real networked systems, such as financial networks, power networks, social networks, protein-protein interaction networks, disease networks, and gene regulatory networks \cite{vinayagam2016controllability, zhang2015determining, liu2014detection, wang2014controllability, ravindran2016controllability, ravindran2017identification, li2016characterizing, wang2018controllability}. In the meanwhile, research on different directions of control were also stimulated, such as edge controllability \cite{nepusz2012controlling, pang2017universal} , exact controllability \cite{yuan2013exact, gao2016emergence}, strong structural controllability \cite{jarczyk2011strong} and dominating sets \cite{nacher2016minimum, ishitsuka2016critical, molnar2014dominating}, which significantly advanced our understanding on this fundamental problem.

However, except for a few works considering the temporal feature of networks \cite{xiao2014edge, hou2015enhancing, wang2012optimizing, wang2016effective, thalmeier2016action}, most of the current advances on network controllability focus on the static network, in which the number of nodes is fixed and connected by a fixed number of links that do not change over time. But real networks are growing, with new links and nodes constantly added to the system \cite{barabasi1999emergence}. To the best of our knowledge, there is no study on the general principle for the change of controllability in growing networks. In this work, we analyze controllability of evolving networks and propose a general rule for the change of driver nodes under network expansion. This rule allows us to further study a problem of network augmentation subject to the controllability constraints \cite{wang2016effective}. The maximum number of new nodes that can be added to the network while keeping the controllability unchanged is difficulty to obtain. However, the upper and lower bound of this problem can be efficiently identified. The upper and lower bound are also affected by different types of degree correlations in directed networks. In the following discussion, we will briefly review the framework of identifying minimum driver nodes and a node classification scheme based on the multiplicity feature in choosing driver nodes. With these basic concepts, we propose a general rule for the change of driver nodes when a new node is added and connected to existing nodes in the network. Finally, we use this rule to solve a problem of maintaining network controllability while adding new nodes to the system.

\section{Results}
\subsection{Network structural controllability and node classification}
A dynamical system is controllable if it can be driven from any initial state to any desired final state within finite time. In many systems, altering the state of a few nodes is sufficient to drive the dynamics of the whole network. For a linear time-invariant system, the minimum driver node set (MDS) can be identified efficiently \cite{liu2011controllability}. First, a directed network is converted into a bipartite graph by splitting a single node in the directed network into two nodes in a bipartite graph, forming two disjoint sets of + and - nodes. Consequently, a directed link from node $i$ to $j$ in the directed network becomes a link from node $i^{+}$ to node $j^{-}$ in the bipartite graph. Then the maximum matching \cite{Hopcroft1971An, Zdeborov2006The, Tao2015An} of the bipartite graph is identified where one node can at most match another node via one link. The the unmatched nodes in the - set are the driver nodes. By imposing properly chosen signals on these $N_D$ driver nodes of the MDS, we can yield control over the whole system.

\begin{figure}[h]
  \centering
  \includegraphics[width=12cm]{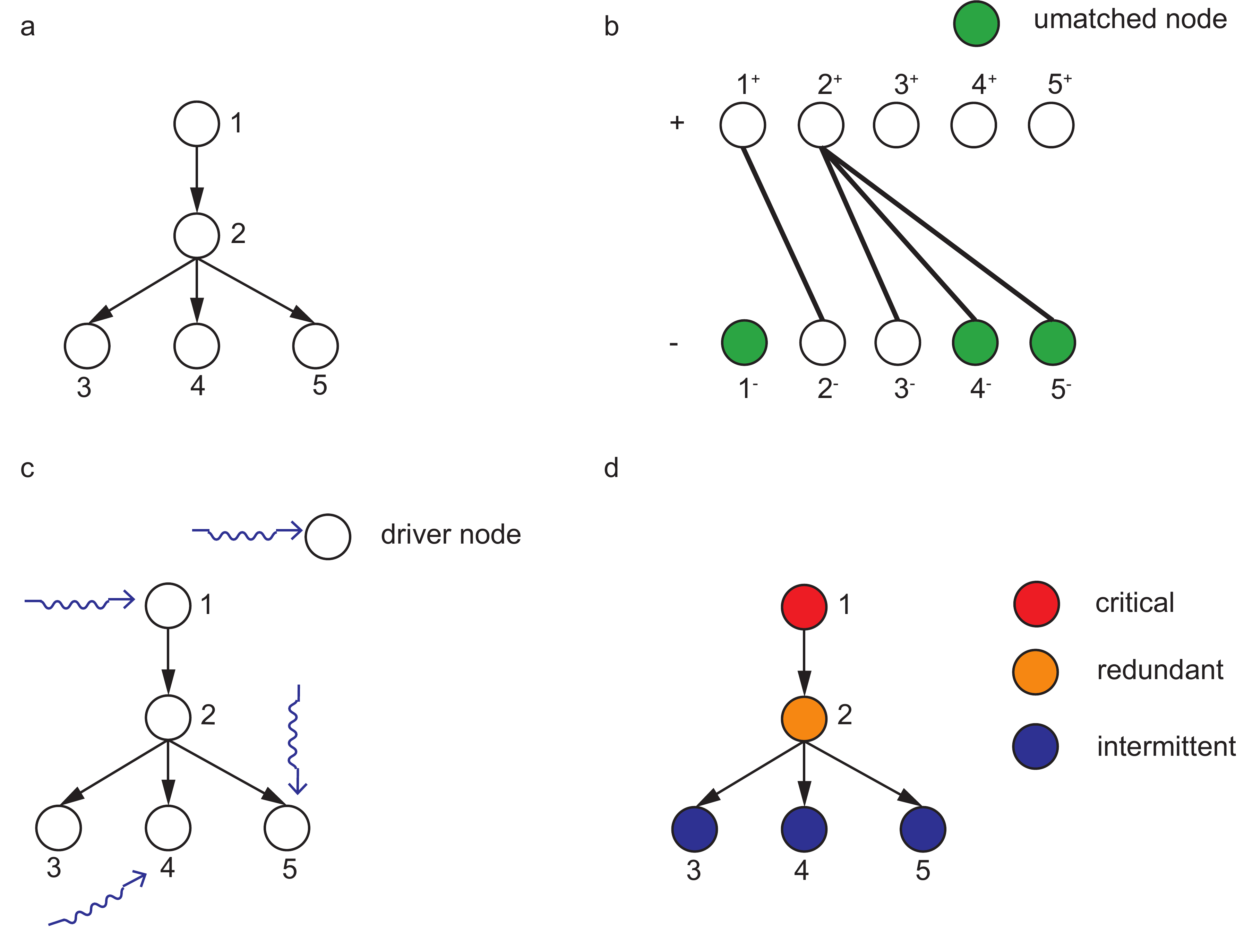}
  \caption{(\textbf{a}) A directed network with five nodes. (\textbf{b}) The directed network in ($\bf a$) can be transferred into a bipartite network by splitting a node to two nodes in the bipartite network. The maximum matching is then performed in the bipartite network, leaving node $1^{-}$, $4^{-}$ and $5^{-}$ unmatched. (\textbf{c}) One minimum driver node set (MDS) is obtained with the maximum matching in ($\bf b$). The whole network is controllable by controlling node 1, 4 and 5. (\textbf{d}) Node classification based on the likelihood of being a driver node. Node 1 is critical, node 2 is redundant and nodes 3, 4, 5 are intermittent. }
 \label{fig.1}
\end{figure}

The number of driver nodes necessary and sufficient for control, $N_D$, is fixed for a given network. However, there are multiple choices of MDSs with the same $N_D$ (Fig. \ref{fig.1}a ), giving raise to a multiplicity feature \cite{jia2013emergence, jia2013control}. Correspondingly, a node classification scheme is proposed based a node's likelihood of being in the MDS. A node may appear in all MDSs. Hence this node is {\it critical} because the network can not be under control without controlling this node. A node may not appear in any MDSs. Consequently this node is {\it redundant} as it does not require any external inputs. The rest kind of nodes that may appear in some but not all MDSs is {\it intermittent}. It has been found that a node is critical if and only if it has no incoming links \cite{jia2013emergence}. Hence, the fraction of critical nodes in the network ($n_c$) is solely determined by the degree distribution, which equals the fraction of nodes with zero in-degree. The redundant nodes in a network can be identified by an algorithm with O($LN$) complexity. The intermittent nodes are therefore readily known once the critical and redundant nodes are identified.

\subsection{Controllability change in growing networks}
An aim of this work to answer the question about how the number of driver nodes $N_D$ changes when a new node is added to the network with new links pointing to / from the existing nodes. For simplicity, we separately analyze two cases when the new node has only incoming links and only outgoing links. Indeed, the correlation between a node's in- and out-degree does not affect the overall controllability\cite{posfai2013effect}. Therefore, the general case when adding a new node with both incoming and outgoing links can be considered as a process that adds one node with only outgoing links and one node with only incoming links, and then merges these two nodes as a single node.

We first consider adding a single node to a network which has only outgoing links. Our conclusion is that if there is one new link connected to a non-redundant node (either a critical node or an intermittent node, denoted by NR node for short) in the original network, the number of drive nodes will stay the same. Otherwise, if all links are connected to redundant nodes (denoted by R nodes for short) in the original network, the number of drive nodes will increase by 1. While we put detailed proof in the Appendix (see Appendix A), this conclusion can be intuitively understood as follows. A node without incoming links always requires an independent external signal to control. Hence when this node is added to a network, the number of driver nodes will either increase by 1 or stay unchanged, depending on whether an existing driver node in the original network would become a non-driver node after adding this new node.  Since a redundant node could never become a driver node, linking to them will not change the original number of external signals. In this case, $N_D$ will increase by 1. In contrast, a critical node is always a driver node, and an intermittent node can become a driver node in some circumstances. Connecting to these two types of nodes can save one original signal, making $N_D$ stay the same.

The the situation that a single node with only incoming links is added to a network can be analyzed in a similar way by introducing the transpose network, in which the direction of all links in the original network is reversed. The value $N_D$ is the same in both the original network and the transpose network. Therefore, the problem that how $N_D$ would change if adding a node with only incoming links is equivalent to the problem that how $N_D$ would change if adding a node with only outgoing links in the transpose network. Correspondingly, we can first identify a node's category in the transpose network, i.e.\ identify whether a node is redundant or non-redundant in the transpose network. Then we can apply the result above and reach a conclusion that if there is one new link from a non-redundant node in the transpose network (denoted by NR$^{T}$ node for short), $N_D$ will stay the same. Otherwise, if all links are from the redundant nodes in the transpose network (denoted by R$^{T}$ nodes for short), $N_D$ will increase by 1.

\begin{figure}[h]
 \centering
  \includegraphics[width=10cm]{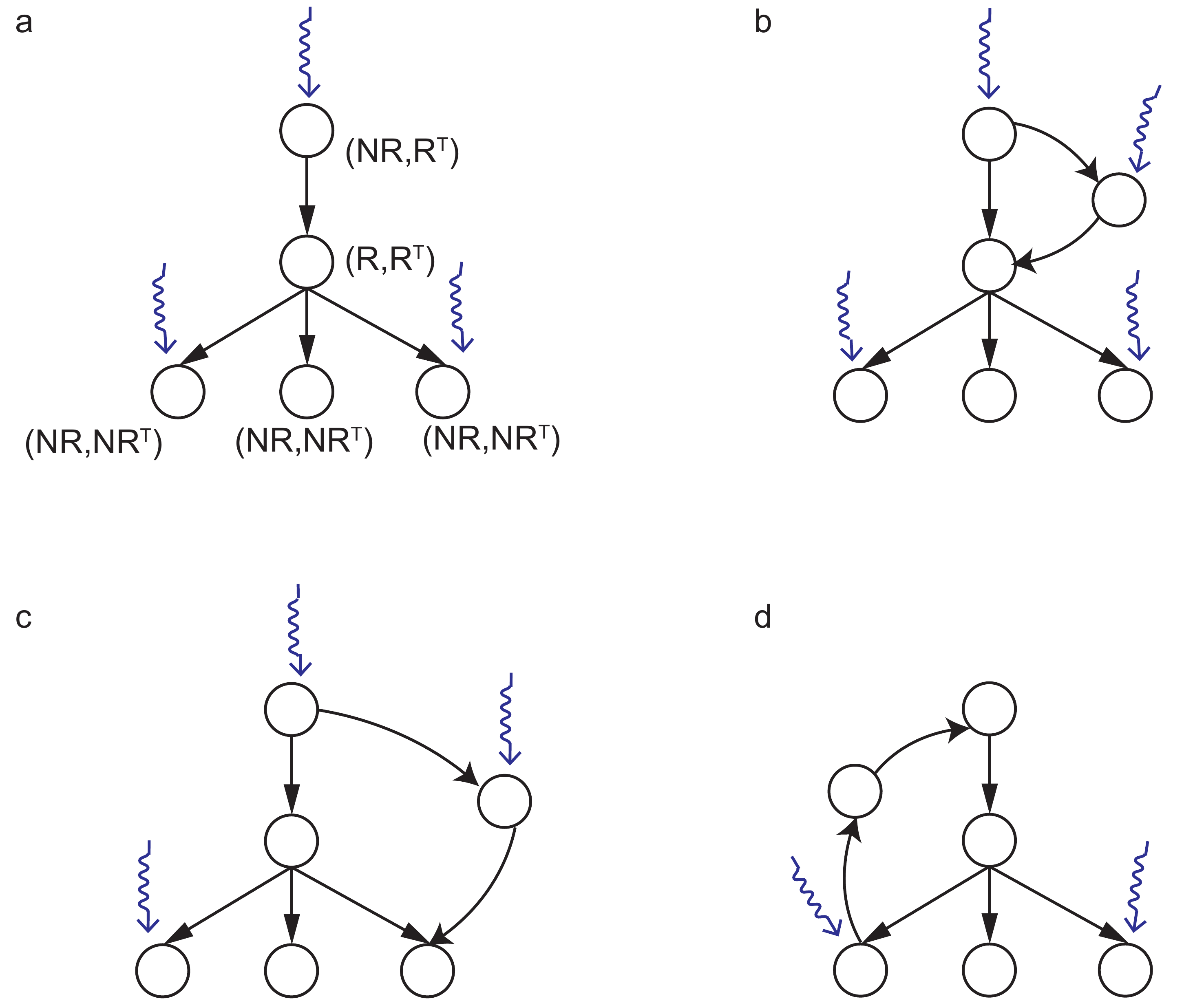}
 \caption{(\textbf{a}) A network with five nodes that can be control via three driver nodes ($N_D$=3). The category of the node, i.e.\ R or NR in the original network, and R$^{T}$ or NR$^{T}$ in the transpose network, is labeled for each of the five nodes. When a new node with one out-going link and one-incoming link is added to the network, (\textbf{b}) $N_D$ increases by 1 when the out-link connects to a R node and the in-link is from a R$^{T}$ node. (\textbf{c}) $N_D$ keeps the same when the out-link connects to a NR node and the in-link is from a R$^{T}$ node. (\textbf{d}) $N_D$ decreases by 1 when the out-link connects to a NR node and the in-link is from a NR$^{T}$ node.}
 \label{fig.2}
\end{figure}

To quickly summarize, when a node with only outgoing links is added, there are two options in the change of $N_D$: (1) $N_D$ increases by 1 if all links are connected to R nodes; (2) $N_D$ keeps the same if at least one link is connected to a NR node. Likewise, when a node with only incoming links is added, there are also two options in the change of $N_D$: (1) $N_D$ increases by 1 if all links are from R$^{T}$ nodes; (2) $N_D$ keeps the same if at least one link is from a NR$^{T}$ node. As mentioned above, the general case when adding a new node with both incoming and outgoing links can be considered as a process that adds one node with only outgoing links and one node with only incoming links, and then merges these two nodes as a single node. It is also noteworthy that $N_D$ will decrease by 1 during this particular merging process (see Appendix B). Taken together, we have the general conclusion on the change of controllability as follows:
\begin{itemize}
\item[-] Identify the category of all existing nodes, i.e.\ R or NR in the original network, and R$^{T}$ or NR$^{T}$ in the transpose network (Fig. \ref{fig.2}a).
\item[-] $N_D$ increases by 1 if all out-links are connected to R nodes and all in-links are from R$^{T}$ nodes (Fig. \ref{fig.2}b).
\item[-] $N_D$ keeps the same if all out-links are connected to R nodes and at least one in-link is from a NR$^{T}$ node, or at least one out-link is connected to a NR node and all in-links are from R$^{T}$ nodes (Fig. \ref{fig.2}c).
\item[-] $N_D$ decreases by 1 if at least one out-link is connected to a NR node and at least one in-link is from a NR$^{T}$ node  (Fig. \ref{fig.2}d).
\end{itemize}

\subsection{A case study of network argumentation problem}

A recent study raises an interesting problem about network argumentation: what is the maximum number of nodes that can be added to a network while keeping $N_D$ unchanged \cite{wang2016effective}. The problem is under several constraints such that some trivial solutions are excluded. First, the new nodes added have only one out-going link. The constraint on one link slightly simplifies the problem. But it also means that the new nodes added are not able to form a cycle. Second, the new nodes are not allowed to connect to critical nodes, i.e.\ the nodes in the network with zero in-degree. Therefore, the trivial solution that new nodes connect one after another to form a directed path is excluded. Finally, the MDS needed to control the original network is recorded and the new nodes are not allowed to connect to any nodes in the original MDS. Indeed, one trivial solution of this problem is to connect the new node to the node in the MDS to keep $N_D$ unchanged. This constraint excludes this trivial solution, and it significantly increases the difficulties of the problem, which will be explained later.

The problem itself has several implications to real systems \cite{wang2016effective,jalili2018effective}, which is not the focus of our work. We are interested in identifying the maximum number of nodes $N_a$ that can be added to the network  in this problem. One intuitive answer is $N_a = N_D - N_c$. The reason is as follows. Because the new node added has zero in-degree, it needs to be always controlled and be the driver node once it is added. To keep $N_D$ the same, we can add at most $N_D$ of these new nodes. Because nodes with zero in-degree are not allowed to be connected, the number of critical nodes in the original network should be deducted, which yields the answer $N_a = N_D - N_c$.

\begin{figure}
  \centering
\includegraphics[width=14cm]{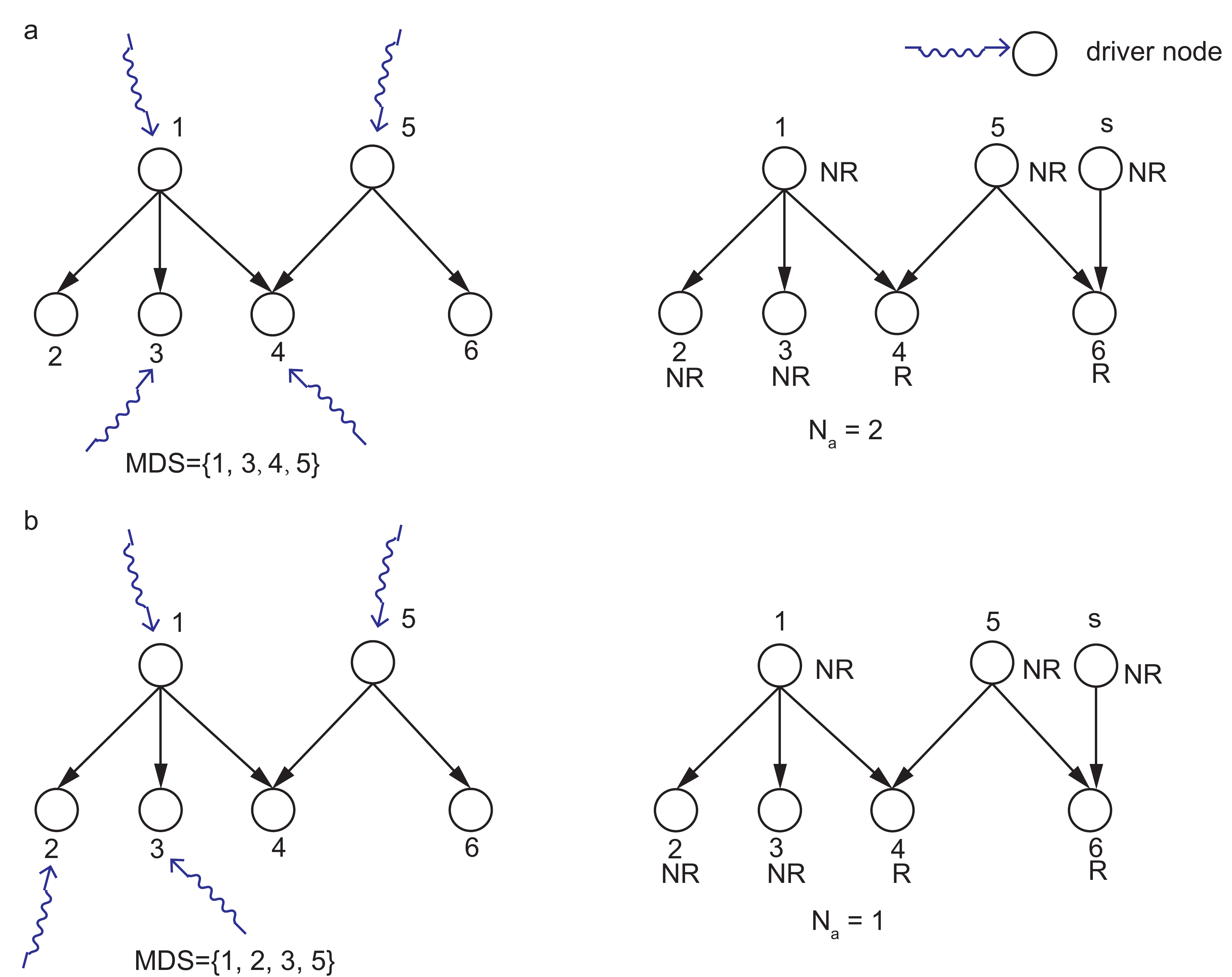}
  \caption{An example of how the original MDS can affect $N_a$. (\textbf{a}) Node 1, 3, 4, 5 form the MDS. After adding a new node s connecting to node 6, node 2 remains to be a NR node, allowing an extra node to be added. In this case, $N_a=2$. (\textbf{b}) Node 1, 2, 3, 5 form the MDS. After adding a new node s connecting to node 6, there is no NR node that is not included in the original MDS. In this case, $N_a=1$.}
  \label{fig.3}
\end{figure}

However, this solution does not effectively satisfy the third constraint. Because the new node can not connect to nodes in the original MDS, the way that the original network is controlled will affect $N_a$. Indeed, in our conclusion of controllability change, we show that the new node has to connect to a NR node to keep $N_D$ the same. But if the NR node is also in the MDS, it is not allowed to be connected. Fig. \ref{fig.3} shows a good example about how the original MDS would affect $N_a$.

\begin{figure}[h]
  \centering
 \includegraphics[width=14cm]{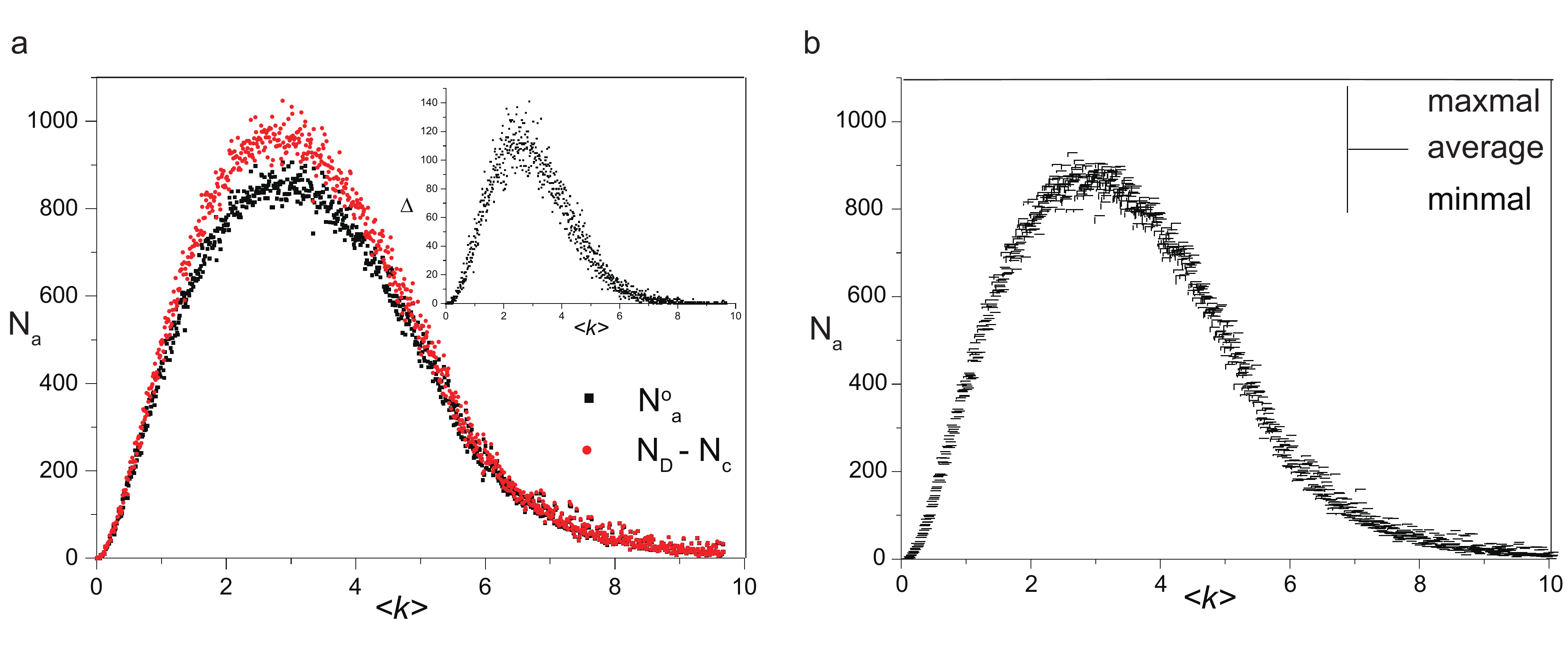}
 \caption{(\textbf{a}) The upper bound ($N_D-N_c$) and the lower bound ($N^o_a$) of $N_a$ for ER network with $N=$10000 and varying $\langle k \rangle$. $N^o_a$ is obtained via one realization of MDS. Both the upper and lower bound vary non-monotonically with $\langle k \rangle$. The difference between the upper and lower bound, $\Delta$ (insert), also varies in a similar trend as those of $N_D-N_c$ and $N^o_a$. (\textbf{b}) The average, maximal and minimal value of $N^o_a$ based on 100 randomly generated MDSs in ER network with $N=$10000 and varying $\langle k \rangle$. }
 \label{fig.4}
\end{figure}

Hence, $N_D$ - $N_c$ should be the upper bound of $N_a$, but $N_a$ in many cases can be less than $N_D$ - $N_c$. The exact value of $N_a$ turns out to be highly non-trivial, related with solving an integer programming. But based on the principle identified, we can use a greedy algorithm to find the local maximum, denoted by $N^o_a$, which represents the lower bound of $N_a$. The idea is to identify a NR node which is not in the MDS and connect the new node to this NR node. The algorithm (See Appendix C) takes O($NL$) complexity to identify a set of $N^o_a$ nodes for a given choice of MDS that are introduced to control the original network. We find that $N^o_a$ identified using our algorithm can be quite less than $N_D$ - $N_c$. Such difference varies non-monotonically with the average degree of the network $\langle k \rangle$ which reaches the peak at a intermediate value of  $\langle k \rangle$  (Fig. \ref{fig.4}a).

$N^o_a$ depends on the particular choice of MDS and there are multiple MDSs for a given network. To take this multiplicity feature into account, we apply the random sampling method \cite{jia2013control} to generate a collection of random MDSs, in which each MDS gives rise to a $N^o_a$ value. We then calculate the mean, maximal and minimal value of $N^o_a$ based on the collection of MDSs (Fig. \ref{fig.4}b). In general, the mean and maximal value of $N^o_a$ are very close. Statistically $N^o_a$ is not significantly affected by the multiple choices of MDSs. But there exist rare cases when the $N^o_a$ value is much less than its mean.

\begin{figure}[h]
 \centering
 \includegraphics[width=14cm]{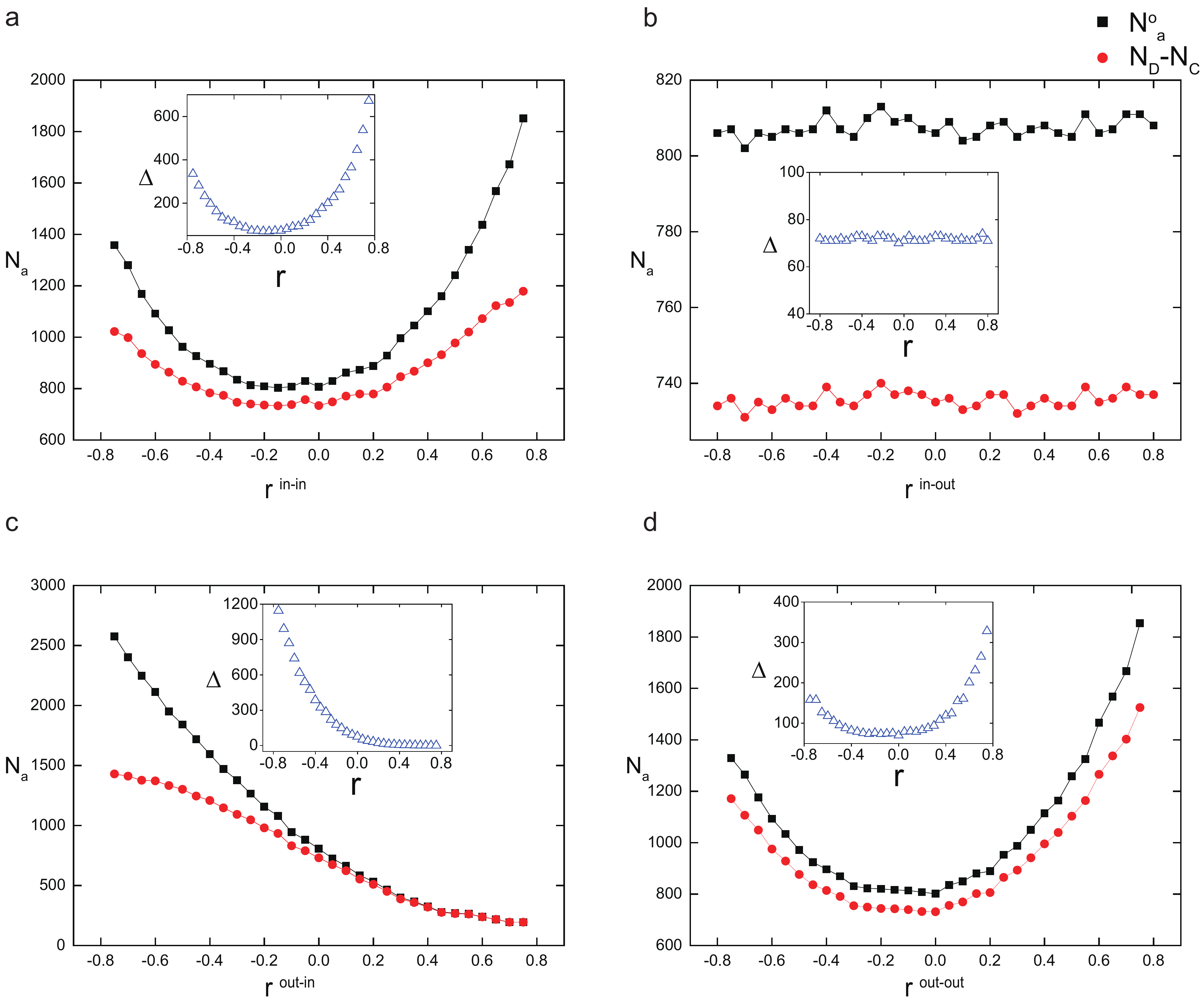}
  \caption{The relationship between the degree correlation and the $N_a$ in ER network with $N=10000$ and $\langle k \rangle = 4$. The four types of degree correlation are denoted by $r^{in-in}, r^{in-out}, r^{out-out}, r^{out-in}$.  The upper bound ($N_D-N_c$), the lower bound ($N^o_a$), and the difference between the two ($\Delta$, in the inset) change very similarly.}
  \label{fig.5}
\end{figure}

Finally, we analyzed the effect of degree correlation on $N_a$. In most real systems, connections between nodes are not neutral \cite{newman2003mixing, pastor2001dynamical, newman2002assortative}. Nodes are with certain tendency to connect to nodes with similar or different degree. Such tendency, or degree correlation, is usually quantified by Pearson correlation coefficient between the degree of two nodes connected by a single link \cite{newman2002assortative, newman2003mixing}. In directed networks, a node is characterized by both in- and out-degree. Hence, there are four different quantification of degree correlation\cite{foster2010edge,posfai2013effect}. More specifically, for a direct link starts at node \emph{s} and ends at node \emph{t},  the degree correlation r is given by:

\begin{equation}\label{eq.1}
  r^{(\alpha,\beta)} = \frac{L^{-1}\sum_{i}\alpha_{i}^{s}\beta_{i}^{t} - [L^{-1}\sum_{i}1/2(\alpha_{i}^{s}+\beta_{i}^{t})]^2}{L^{-1}\sum_{i}1/2(\alpha_{i}^{s^{2}} + \beta_{i}^{t^2}) - [L^{-1}\sum_{i}1/2(\alpha_{i}^{s} + \beta_{i}^{t})]^2},
\end{equation}
where $L$ is the total number of links in the network, $\alpha$, $\beta$ $\in \{in, out\}$ corresponds to the two different types of degree. The four types of degree correlation are hence denoted by $r^{in-in}, r^{in-out}, r^{out-out}, r^{out-in}$.

Since $N_c$ does not change with degree correlation but only depends on the number of nodes with 0 in-degree ($P_\text{in}(0)$), the upper bound of $N_a$, which is $N_D - N_c$, changes with $N_D$ alone. It does not change with $r^{in-out}$, increases with the absolute value of $r^{in-in}$ and $r^{out-out}$, and monotonically decreases with $r^{out-in}$ \cite{posfai2013effect} (Fig. \ref{fig.5}). The lower bound $N^o_a$, identified using our method, follows a similar trend of $N_D$ in all cases. Furthermore, the difference $\Delta$ between the upper and lower bound also shows a similar trend as that of $N_a$ and $N^o_a$.

\section{Discussion}

In summary, we study the change of network controllability in growing networks. We introduce two sets of node categories, R and NR, and R$^{T}$ and NR$^{T}$. We find that the number of driver nodes $N_D$ can increase by 1, stay the same or decrease by 1 when a new node is added. The change relies on the categories of nodes (R or NR) that the out-going links connects to and the categories of nodes (R$^{T}$ or NR$^{T}$) that the in-coming links are from. This principle on the change of controllability helps us to solve a recently proposed problem on network augmentation, the maximum number of nodes $N_a$ that can be added to a network while keeping $N_D$ unchanged. We propose an algorithm that can efficiently finds the lower bound of $N_a$. We demonstrate how the upper bound and lower bound change with average degree and the degree correlation of the network.

The results presented have many potential applications in future works \cite{wang2016geometrical}. Network expansion or augmentation is a ubiquitous feature in our rapidly growing technological society such as adding nodes or edges to an existing network. Generally, when the network is going to be larger, there will be more nodes required to achieve full control and also the cost of control the network will increase. Our approach can offer insights for future work exploring the augmentation of nodes in control and offer fundamental tools to explore control in temporal complex systems.

\section*{Acknowledgement}
This work is supported by the Natural Science Foundation of China (No. 6160309). M. C. is also supported by the Nature Science Foundation of Jiangsu Province (No. BK20150344), China Postdoctoral Science Foundation (No. 2016M601885).

\appendix

\section{}

Our conclusion says that if there is one new link connected to a non-redundant node (either a critical node or an intermittent node, denoted by NR node for short) in the original network, the number of drive nodes will stay the same. Otherwise, if all links are connected to redundant nodes (denoted by R nodes for short) in the original network, the number of drive nodes will increase by 1. The proof of this conclusion is best described in a bipartite graph. Therefore, we will change the terminology from the ``driver node in a directed network'' to the ``matched or unmatched node in the - set of a bipartite graph''. In particular there are several equivalent terms. \\
number of driver nodes $=$ number of nodes in a set (either $+$ or $-$) $-$ number of matched pairs (or number of matched nodes in a set)  \\
redundant node $=$ always matched node in the - set of a bipartite graph,  i.e.\ the node is matched in all different maximum matching configurations \\
non-redundant node $=$ not always matched node in the - set of a bipartite graph. This includes the node that is not matched in the current maximum matching configuration, and the node that is currently matched but can be unmatched in a different maximum matching configuration\\

\begin{figure}
 \centering

 \includegraphics[width=10cm]{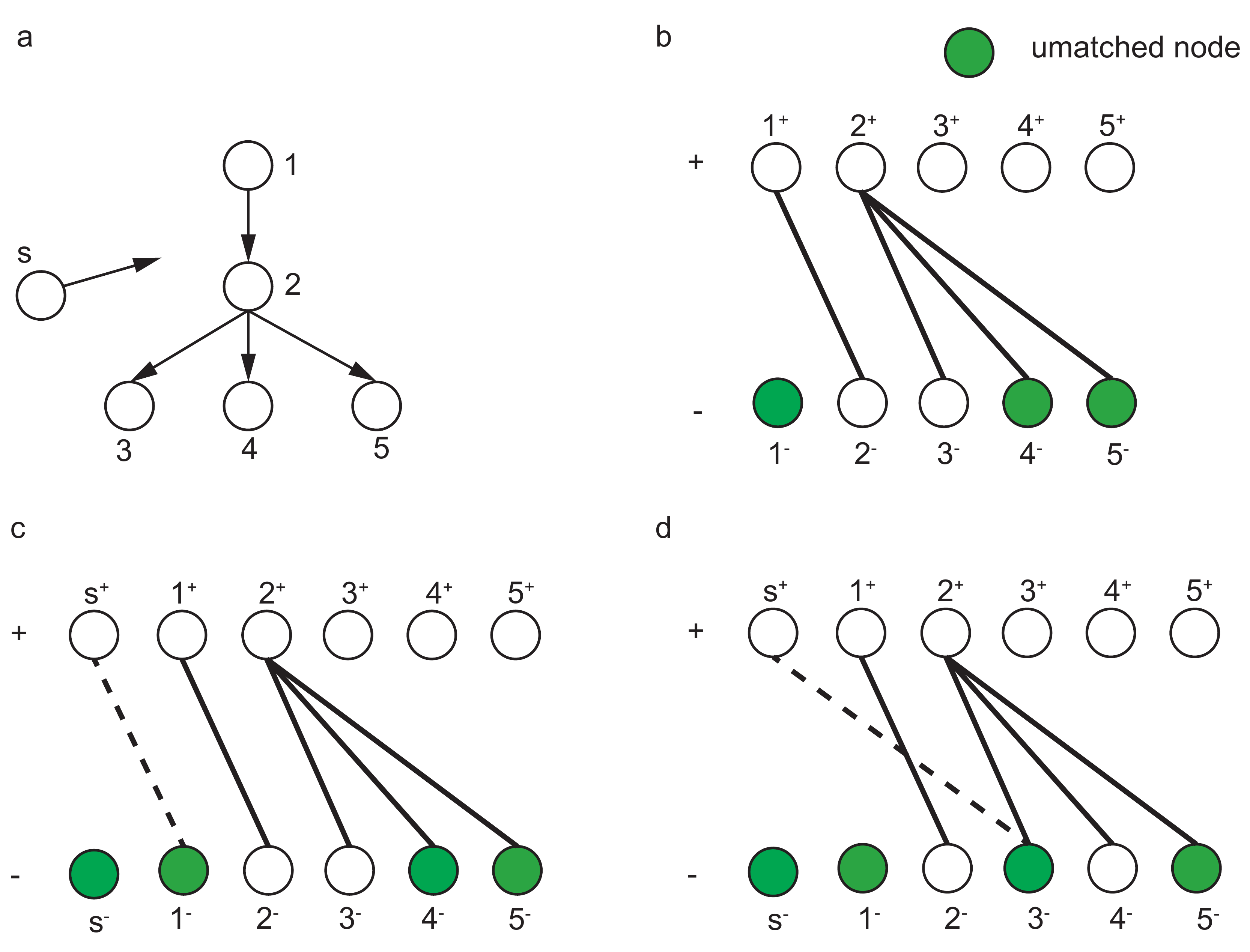}
  \caption{(\textbf{a}) A new node s with one out-going link is added to a directed network with five nodes. (\textbf{b}) The corresponding bipartite network with a maximum matching obtained. Node $1^-$, $4^-$ and $5^-$ are unmatched in the original network. (\textbf{c}) Node $1^-$ is unmatched. If node $s^-$ connects to node $1^-$, a new maximum matching is achieved with three matched pairs. (\textbf{d}) Node $3^-$ is currently matched, but it is not always matched. Consequently, there is a different matching configuration that preserves the same number of matched pairs while leaving node $3^-$ unmatched. Therefore, when node $s^-$ connects to node $3^-$, a new maximum matching is with three matched pairs can be achieved.}
  \label{fig.6}
\end{figure}

Now let us consider the case when a new node $s$ with one out-going link is added to a directed network (Fig. \ref{fig.6}a). In the bipartite graph representation, this is to add a node $s^{-}$ with zero link and a node $s^{+}$ with one link. Assuming that the node $s^{+}$ connects to is $t^{-}$, there are 3 different situations:
\begin{enumerate}
\item Node $t^{-}$ is unmatched. Then a new maximum matching is achieved by matching node $s^{+}$ and node $t^{-}$ (Fig. \ref{fig.6}b).
\item Node $t^{-}$ is matched in the current maximum matching configuration, but it is not always matched. Because node $t^{-}$ is not always matched, there are configurations that node $t^{-}$ is unmatched but yields the same number of matched pairs. Then we can always change the matching configuration to such that node $t^{-}$ is unmatched and them match the pair node $s^{+}$ and node $t^{-}$. In the terminology of Hopcroft-Karp algorithm that is applied to find the maximum matching in bipartite graph \cite{Hopcroft1971An}, this means that there exists an augmentation path that can go through node $s^{+}$ and node $t^{-}$.
\item Node $t^{-}$ is matched and is always matched. In this case, there is no maximum matching configuration with $t^{-}$ unmatched. Hence, node $s^{+}$ and node $t^{-}$ can not be matched. In other words, there is no augmentation path that can go through node $s^{+}$ and node $t^{-}$.
\end{enumerate}

In case 1 and 2 (connecting to a node that is not always matched), the number of maximum matching increases by 1, which offsets the increase of total number of nodes. Consequently, the number of drive nodes stays the same. In case 3 (connecting to an always matched node), the number of maximum matching stays the same but the number of nodes increases by 1. Hence, the number of drive nodes will increase by 1.

\section{}

Assume that node $s_1$ is with only in-coming links and node $s_2$ is with only out-going link. In the corresponding bipartite network, node $s_1^{+}$ and node $s_2^{-}$ are with zero degree. Now consider the case that node $s_1$ and node $s_2$ are merged together to form a node $s$ with both in-coming and out-going links. In the bipartite network, it corresponds to the process that node $s_1^{+}$ and node $s_2^{+}$, node $s_1^{-}$ and node $s_2^{-}$ are merged together. Note that node $s_1^{+}$ and node $s_2^{-}$ can not be matched in the original bipartite network. Therefore, the merging does not change the number of matched pairs. But the number of nodes in each set is reduced by 1. Hence, $N_D$ will decrease by 1 in this merging process as $N_D=$ number of nodes in a set $-$ number of matched pairs (see Appendix A).

\section{}

When adding a node with one out-going link, $N_D$ keeps the same if the new node connects to a NR node. To satisfy other constraints, the connected node should not belong to the original MDS (the nodes with zero in-degree in the original network are also in the MDS). The MDS corresponds to a set of - nodes that are not matched in the bipartite network. Therefore, the connected node should be matched but not always matched. Correspondingly, the algorithm is as follows:\\
1. From the original MDS, identify a set $M$ of matched nodes in the - set of the bipartite network. Build the matching configuration corresponds to the set of matched nodes.\\
2. Pick an element in $M$ (denoted by node $i^{-}$).\\
3. Check if node $i^{-}$ is always matched by un-matching this node and check if an alternative matching configuration exists that preserves the number of matched pairs. This is equivalent to check if an argumentation path exist that starts with the node that matches node $i^{-}$.\\
4. If node is $i^{-}$ is always matched, repeat step 2. Otherwise, node $i$ is a NR node and the new node should connect to it.\\
5. Update the matching configuration after adding the new node. Repeat step 2 to identify another NR node to connect.

Note that the maximum matching will change when the new node is added. A NR node can become R node. Therefore, we need to test each node one by one. The complexity to find one maximum matching is O($N^{0.5}$ $L$). A node in the $M$ requires a breadth first search (O($L$) complexity) to test if it is not always matched. The complexity to update the maximum matching after adding a new node is O($L$). The number of nodes in $M$ is proportional to $N$. Therefore, the complexity to find $N_a^o$ is O($NL$).

\bibliographystyle{elsarticle-num} 
\bibliography{reference}

\end{document}